%% file: gro_j1008.tex
\documentclass[iop]{emulateapj}

\usepackage[svgnames,pdftex,hyperref,table]{xcolor}
\usepackage[pdftex]{hyperref}
\hypersetup{colorlinks=True,linkcolor=Black,citecolor=Black,
urlcolor=Black,pagebackref,pdftitle={A Search for Cyclotron Lines in the
2012 Giant Outburst of GRO J1008$-$57},pdfauthor={Eric Bellm}}

\usepackage{amssymb}
\usepackage{nicefrac}

\newcommand{\spm}[2]{\ensuremath{^{+#1}_{-#2}}}

\newcommand{\nustar}{\textit{NuSTAR}\ }
\newcommand{\swift}{\textit{Swift}\ }
\newcommand{\suzaku}{\textit{Suzaku}\ }
\newcommand{\nustarn}{\textit{NuSTAR}}
\newcommand{\swiftn}{\textit{Swift}}
\newcommand{\suzakun}{\textit{Suzaku}}
\newcommand{\gro}{GRO\,J1008$-$57\ }
\newcommand{\gron}{GRO\,J1008$-$57}

\slugcomment{Submitted to the Astrophysical Journal}

\shorttitle{Observations of GRO\,J1008$-$57}
\shortauthors{Bellm et al.}

\begin{document}


\title{Confirmation of a High Magnetic Field in
GRO\,J1008$-$57}


\author{Eric C. Bellm\altaffilmark{1},
Felix F{\"u}rst\altaffilmark{1},
Katja Pottschmidt\altaffilmark{2,3},
John A. Tomsick\altaffilmark{4},
Steven E. Boggs\altaffilmark{4},
Deepto Chakrabarty\altaffilmark{5},
Finn E. Christensen\altaffilmark{6},
William W. Craig\altaffilmark{4,7},
Charles J. Hailey\altaffilmark{8},
Fiona A. Harrison\altaffilmark{1},
Daniel Stern\altaffilmark{9},
Dominic J. Walton\altaffilmark{1}, 
J\"orn Wilms\altaffilmark{10},
and William W. Zhang\altaffilmark{11}
}

\altaffiltext{1}{Cahill Center for Astronomy and Astrophysics, California
Institute of Technology, Pasadena, CA 91125; ebellm@caltech.edu.  }
\altaffiltext{2}{Center for Space Science and Technology, University of
Maryland
Baltimore County, Baltimore, MD 21250, USA}
\altaffiltext{3}{CRESST and NASA Goddard Space Flight Center, Astrophysics
Science
Division, Code 661, Greenbelt, MD 20771, USA}
\altaffiltext{4}{Space Sciences Laboratory, University of California,
Berkeley, CA 94720}
\altaffiltext{5}{Kavli Institute for Astrophysics and Space Research, 
Massachusetts Institute of Technology, Cambridge, MA 02139}
\altaffiltext{6}{DTU Space - National Space Institute, Technical
University of Denmark, Elektrovej 327, 2800 Lyngby, Denmark}
\altaffiltext{7}{Lawrence Livermore National Laboratory, Livermore, CA 94550}
\altaffiltext{8}{Columbia Astrophysics Laboratory, Columbia University, New
York, NY 10027}
\altaffiltext{9}{Jet Propulsion Laboratory, California Institute of
Technology, Pasadena, CA 91109}
\altaffiltext{10}{Dr. Karl-Remeis-Sternwarte and ECAP, Sternwartstr. 7,
96049 Bamberg, Germany}
\altaffiltext{11}{NASA Goddard Space Flight Center, Greenbelt, MD 20771}







\begin{abstract}

\gro is a high-mass X-ray binary for which several claims of a cyclotron
resonance scattering feature near 80\,keV have been reported.
We use \nustarn, \suzakun, and \swift data from its giant
outburst of November 2012 to confirm the existence of the 80\,keV feature
and perform the most 
sensitive search to date for cyclotron scattering features at lower
energies.  We find evidence for a 78\spm{3}{2}\,keV line in the
\nustar and \suzaku data at $> 4\sigma$ significance, confirming the
detection using \suzaku alone by \citet{tmp_Yamamoto:14:GROCyc}.
A search of both the
time-integrated and phase-resolved data rules out a fundamental at lower
energies with optical depth larger than 5\% of the 78\,keV line. 
These results indicate that \gro has a magnetic field of 
$6.7\times10^{12} (1+z)$\,G,
the highest of known accreting pulsars.

\end{abstract}


\keywords{pulsars: individual (GRO J1008–57); stars: neutron; X-rays: binaries}


\section{Introduction}

\gro is a transient high-mass X-ray binary (HMXB) system with a neutron
star primary and a Be companion. It was discovered by the Burst and
Transient Source Experiment aboard the \textit{Compton Gamma-Ray
Observatory} during a 1.4 Crab giant outburst in July 1993
\citep{Stollberg:93:GROJ1008Discovery,Wilson:1994:GROJ1008Discovery}.
Optical followup identified its Be-type companion and suggested a distance
to the source of 5\,kpc \citep{Coe:94:GROJ1008OptCounterpart}.  

Like other Be/X-ray binaries (Be/XRBs), 
\gro exhibits regular outbursts (Type I) due to accretion
transfers during periastron passages as well as irregular giant (Type II)
outbursts
\citep[for a recent review of Be/XRB systems, see][]{Reig:2011:BeXRBReview}.  
Its Type I outbursts occur predictably at the 249.48 day orbital period
\citep{Kuhnel:2013:GROJ1008,Levine:06:RXTEPeriodicities}.
\citet{Kuhnel:2013:GROJ1008} found that the spectra of \gro during Type I
outbursts are similarly
regular: the continuum spectrum consists of an exponentially cutoff power-law 
and a low-energy black body
component whose properties correlate strongly with source flux.

Accreting pulsars, of which Be/XRBs are a subclass,
characteristically exhibit cyclotron resonant scattering
features (CRSFs) in the hard X-ray band 
due to Compton scattering off of electrons with
orbits quantized by the $\sim10^{12}$\,G 
magnetic field of the neutron star.
The observed line energy provides a direct probe of the magnetic field
strength, with $E_{\rm cyc} = 11.6 B_{12}/(1+z)$~keV, where $B_{12}$ is
the magnetic field strength in units of 10$^{12}$\,G and $z$ is the
gravitational redshift at the emission radius \citep{Canuto:77:CRSF12B12}. 

Based on \textit{CGRO}/OSSE spectra, \citet{Grove:95:GROJ1008CRSF} and \citet{Shrader:99:GROJ1008CRSF} each reported
indications for a possible CRSF at $\sim$88\,keV at low significance
($\sim2\sigma$) for \gron.  Their data 
did not provide energy coverage below 50\,keV
to search for a lower-energy fundamental CRSF at $\sim45$\,keV.  If the
88\,keV feature were confirmed as the fundamental, it would imply that 
\gro has a
magnetic field strength near 10$^{13}$\,G, the highest of any known accreting
pulsar\footnote{\citet{LaBarbera:01:LMCX4Cyc} reported an extremely broad
CRSF centered at 100\,keV for LMC X-4, but these measurements were not
confirmed by \textit{INTEGRAL} \citep{Tsygankov:05:LMCX4NoCyc}.} \citep[e.g.,][]{Caballero:12:XRayPulsarReview}.

Subsequent modeling of data taken over a broader energy band with
\textit{RXTE}, \textit{INTEGRAL}, and \suzaku did not reveal 
a lower-energy fundamental line in the 40--50\,keV region
\citep{Coe:07:GROJ1008Disk,Kuhnel:2013:GROJ1008}, and
detection of the 88\,keV CRSF remained marginal.
\citet{tmp_Wang:14:GROJ1008_INTEGRAL} reported a $\sim3\sigma$ detection
of a CRSF at 74\,keV in a 2009 outburst with \textit{INTEGRAL}.

The regular \gro Type I outburst of September 2012 was followed by several
months of irregular flaring before the source brightened into a 
giant outburst in November 2012.  The increased flux triggered
\textit{MAXI} on November 9
\citep{Nakajima:2012:GROJ1008Brightening} and \textit{Swift}-BAT on
November 13 \citep{Krimm:2012:GROJ1008Brightening}.  Peak flux levels reached
1 Crab in the next week, providing an opportunity to obtain high-statistics
observations of the system in outburst.  \suzaku executed a
Target-of-Opportunity (ToO) observation on November 20 and reported
a detection of a cyclotron line at $E_{\rm cyc} =$74--80\,keV, with the
exact energy depending on the continuum modeling
\citep{Yamamoto:2012:GROSuzakuCyc,tmp_Yamamoto:14:GROCyc}.

\begin{figure}
\includegraphics[width=\columnwidth]{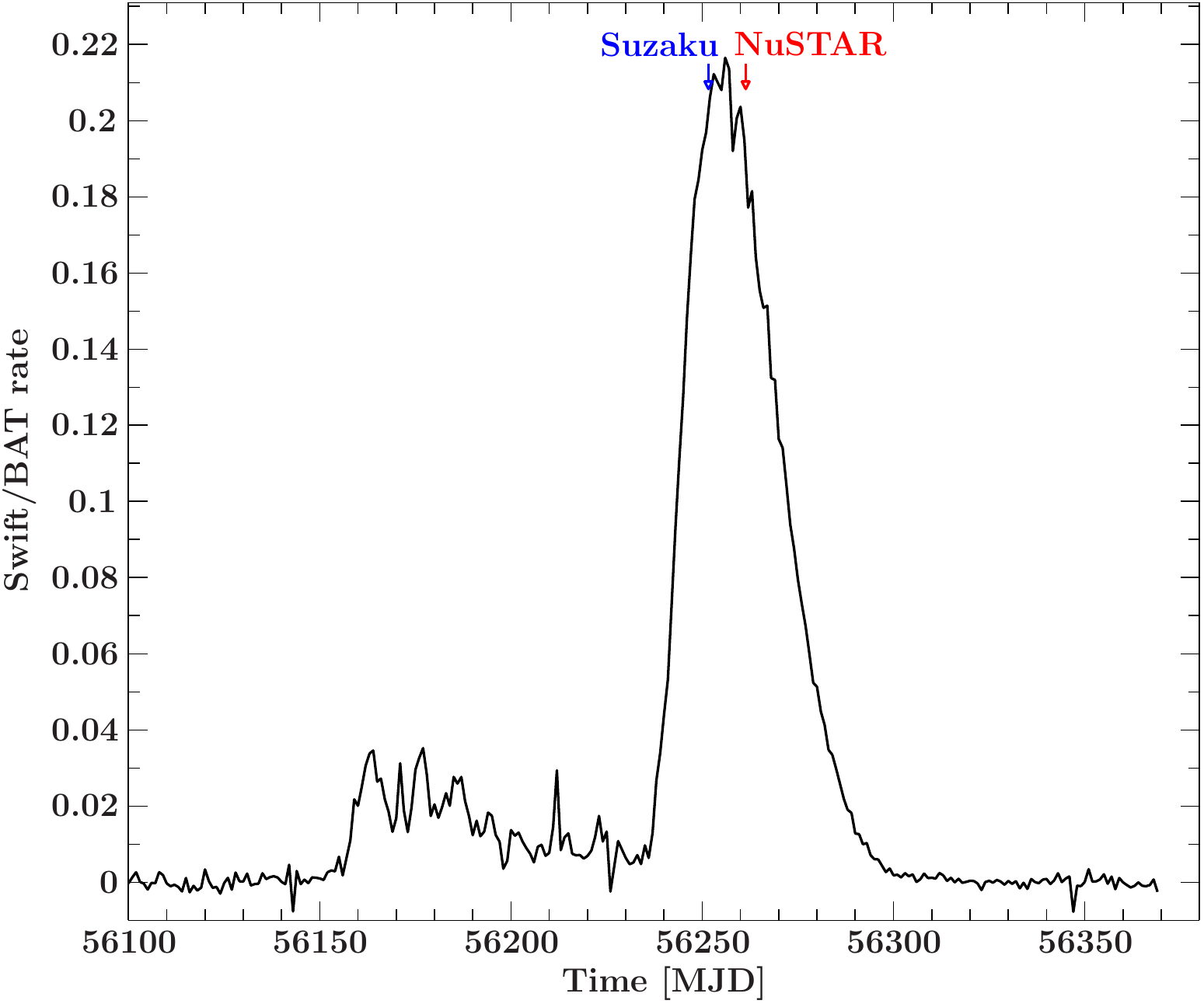}
\caption{
\swiftn-BAT light curve of the giant outburst of \gro with the \nustar and
\suzaku observation times marked. The BAT count rate is in units of 
counts\,cm$^{-2}$\,s${-1}$.
\label{fig:lc}}
\end{figure}

Thanks to its focusing hard X-ray telescopes,
\nustar \citep{Harrison:2013:NuSTAR} provides unprecedented sensitivity
in the broad 3--79\,keV band.  \textit{NuSTAR}'s continuous energy coverage 
removes a major source of systematic errors when fitting broad-band models, 
while the large effective area and lack of pile-up 
enables high-statistics time-resolved 
spectroscopy for bright sources.  \nustar is capable of executing
ToO observations within 24\,hours of trigger and is thus an ideal
instrument with which to study cyclotron lines across a wide range of
magnetic field strengths in neutron star binary systems
\citep[e.g.,][]{Fuerst:13:HerX1,Fuerst:14:VelaX1}.  \nustar observed \gro
on November 20, shortly after the peak of the outburst (Figure
\ref{fig:lc}).

In this paper we combine \nustarn, \swift \citep{Gehrels:04:Swift}, and
\suzaku \citep{Mitsuda:07:Suzaku} observations of the November 2012 giant
outburst in order to obtain the best constraints on the existence
of the putative cyclotron line. \S\ref{sec:observations} describes the
observations and data reduction.  In \S\ref{sec:fits}, we perform a series
of spectral fits of the \nustarn, \suzakun, and \swift data.  We fit
continuum models (\S\ref{sec:cont_fits}) as well as the previously reported
CRSF (\S\ref{sec:cyc_line}) to the data.  Monte Carlo tests confirm the
significance of the feature.  We perform searches for generic CRSFs at
lower energies in both the time-integrated (\S\ref{sec:fundamental_search})
and phase-resolved data (\S\ref{sec:phase_resolved}).  We conclude in
\S\ref{sec:discussion}.

\section{Observations} \label{sec:observations}

\nustar performed a TOO observation of \gro beginning at
UTC 2012-11-30 8:41:07 and ending at UTC 2012-11-30 17:31:07.
The total on-source observation time was 12.4~ksec after excluding
occultation intervals and South Atlantic Anomaly passages.  

We processed the data with HEASOFT 6.15 and 
the \nustar Data Analysis Software (NuSTARDAS) v.\,1.3.0 using CALDB
version 20131223.  We extracted source counts from circular regions with 
4.5\,arcmin radius from both \nustar modules.  Because of the brightness of the
source, flux from the PSF wings was present over most of the focal plane,
preventing extraction of a representative background region.  Instead, we
scaled the background observed during deep pointings on the
Extended \textit{Chandra} Deep Field South region
obtained immediately after the \gro
observations \citep[e.g.,][]{DelMoro:14:NuSTARECDFS}.
The background was selected from the \nustar orbital phases matching the \gro
observation and was extracted from the same detector region as the source.
The background is negligible over most of the \nustar band; it only reaches
10\% of the source count rate at 60\,keV, and is 30--60\% of the source
rate in the 70--78\,keV range.

\swift obtained a 2.3\,ksec snapshot of \gro during the \nustar observation
beginning at UTC 2012-11-30 11:09:25.  
We reduced the \swiftn-X-Ray Telescope \citep[XRT;][]{Burrows:05:XRT}
Windowed Timing mode data using standard
procedures in HEASOFT 6.13 and CALDB version 20120830.  

\suzaku observed \gro earlier in its outburst beginning at 
UTC 2012-11-20 14:44:31; see \citet{tmp_Yamamoto:14:GROCyc} for an
independent analysis of these data. 
The exposure time was 50.4\,ksec with 
the Hard X-ray Detector 
\citep[HXD;][]{Takahashi:07:HXD}.
The X-ray Imaging Spectrometer \citep[XIS;][]{Koyama:07:XIS}
observed the source in burst mode, resulting in an exposure
time of 9.1\,ksec.

We reduced data from XIS
modules 0, 1, and 3 using standard procedures in 
HEASOFT 6.13 and CALDB version 20130724. 
Response files were created using the FTOOL task
\texttt{xisresp} with the medium option, selecting the default binning. 
We used extraction regions with
80 arcsec radius and excluded the inner parts of the PSF, roughly following the
5\% pile-up contours. Pile-up was estimated using the \texttt{pileest}
routine, after correcting for residual attitude wobble using
\texttt{aeattcor2}. We combined the $3\times3$ and $5\times5$ editing modes
where available into one spectrum using \texttt{XSELECT}.

We reduced data from the HXD
with the standard pipeline using calibration files as published
with HXD CALDB 20110913. Spectra were extracted using the tools
\texttt{hxdpinxbpi}
and \texttt{hxdgsoxbpi} for the PIN diodes and GSO scintillator, 
respectively. We obtained the tuned
background models from the \suzaku
website\footnote{\url{ftp://legacy.gsfc.nasa.gov/suzaku/data/background/pinnxb_ver2.0_tuned/}
and \url{ftp://legacy.gsfc.nasa.gov/suzaku/data/background/gsonxb_ver2.6/
}}, as well as the recommended
additional ARF for the GSO.

\section{Spectral Fitting} \label{sec:fits}

We fit the data using the \textit{Interactive Spectral Interpretation
System} \citep[ISIS;][]{citeisis_joern}
v1.6.2-19.  For all instruments except for the
\suzaku GSO data (for which the binning scheme was determined by the
background modeling), we rebinned the data to $\sim$\nicefrac{1}{3} of the
FWHM of the energy resolution to avoid oversampling the intrinsic detector
resolution.  We minimized $\chi^2$ in our fits to the data.

The high source flux highlights systematic uncertainties in the response
matrices, so we exclude some regions from spectral fits.
We fit the \nustar data in the 5--78\,keV range.  The \nustar response
falls off sharply beginning around 78\,keV, so this upper bound minimizes the
effect of response modeling 
uncertainties on our cyclotron line fits.  The \nustar
data showed residual deviations in the 3--5\,keV range when fit with data from
\swift and \suzakun, so due to the unusual brightness of the source 
we omit this region to avoid biasing the fit.
We also omit the \nustar data from 68--70\,keV, which is near the
tungsten K-edge and has a known response feature which could bias our
cyclotron line searches.
Similarly, we omit the \swift data in the 0.2--1\,keV range and above 9\,keV
due to residual
features not seen in the XIS data.  We also apply a 3\% systematic error
per spectral bin.  Finally, we fit the \suzaku XIS data in
the bands suggested by \citet{Nowak:11:CygX1}: 0.8--1.72\,keV,
1.88--2.19\,keV, and 2.37--7.5\,keV.  We fit the PIN data in the
20--70\,keV band and GRO in the 60--120\,keV band.

\subsection{Continuum Fitting} \label{sec:cont_fits}

We fit two models frequently used in modeling accreting pulsar
spectra to the time-integrated continuum spectra:
a powerlaw with a high-energy cutoff, and an \texttt{npex} model consisting
of two powerlaws with negative and positive spectral indices and an
exponential cutoff \citep{Makishima:99:NPEX}.  
We also included a Gaussian iron line and a low-energy black body component.  
For fits including data from \suzaku XIS, a second Gaussian component was
needed to adequately fit the iron line complex.
We used an updated version of the \citet{Wilms:2000:wilmAbundances}
absorption model (\texttt{tbnew}) as a neutral absorber 
with \texttt{wilm} abundances \citep{Wilms:2000:wilmAbundances} and
\texttt{vern} cross-sections \citep{Verner:96:vernCrossSections}.
For the powerlaw with high-energy cutoff, we
removed residuals due to the discontinuity at the cutoff energy with a
Gaussian absorber tied to the cutoff energy \citep[e.g.,][and references
therein]{Coburn:02:AccretingPulsars}.  We allowed the normalizations to
vary between all instruments being fit.

\begin{figure}
\includegraphics[width=\columnwidth]{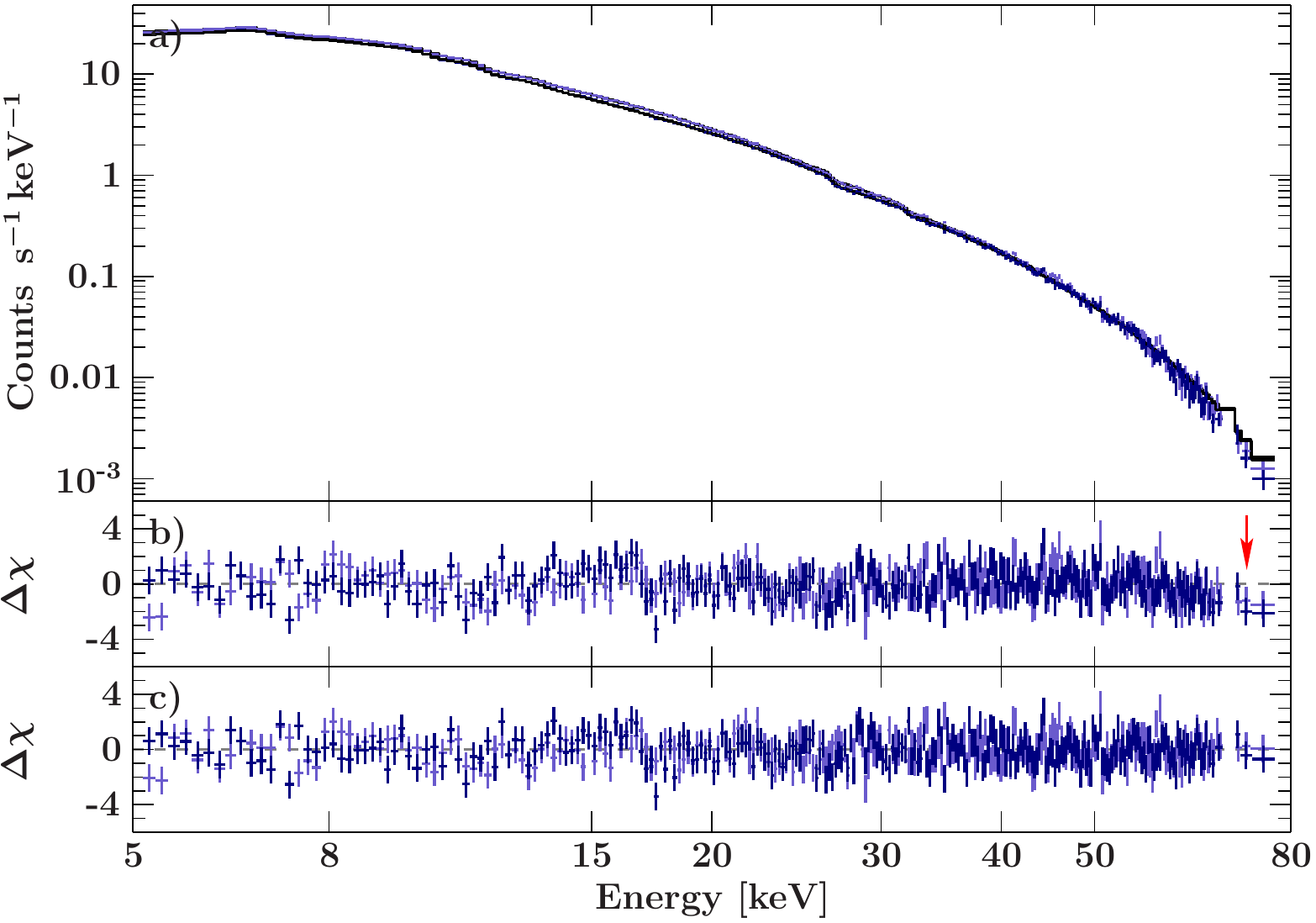}
\caption{
Panel a) Count spectrum and \texttt{npex} model fit to the \nustar data.  
Panel b) Residual plot for the \texttt{npex} fit.
Panel c) Residual plot for a \texttt{npex} fit with \texttt{cyclabs}
component.
An arrow in panel b shows the centroid of the CRSF fit in panel c.
\label{fig:spectra_n}}
\end{figure}

\begin{figure}
\includegraphics[width=\columnwidth]{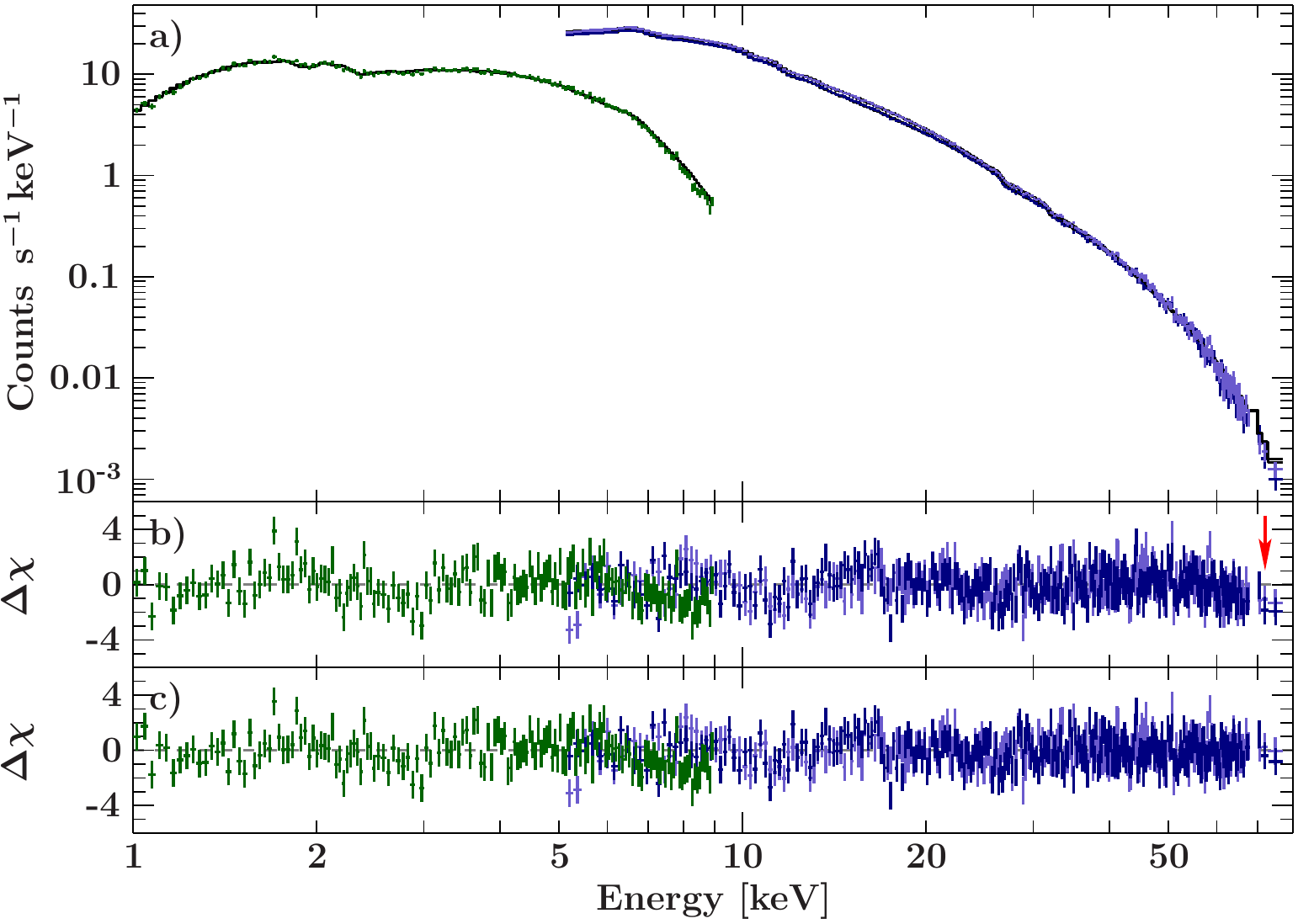}
\caption{
Joint \nustarn--\swiftn-XRT fit.  \nustar data are in blue, XRT data are in
green.  Panels as in Figure \ref{fig:spectra_n}.
\label{fig:spectra_nw}}
\end{figure}

\begin{figure}
\includegraphics[width=\columnwidth]{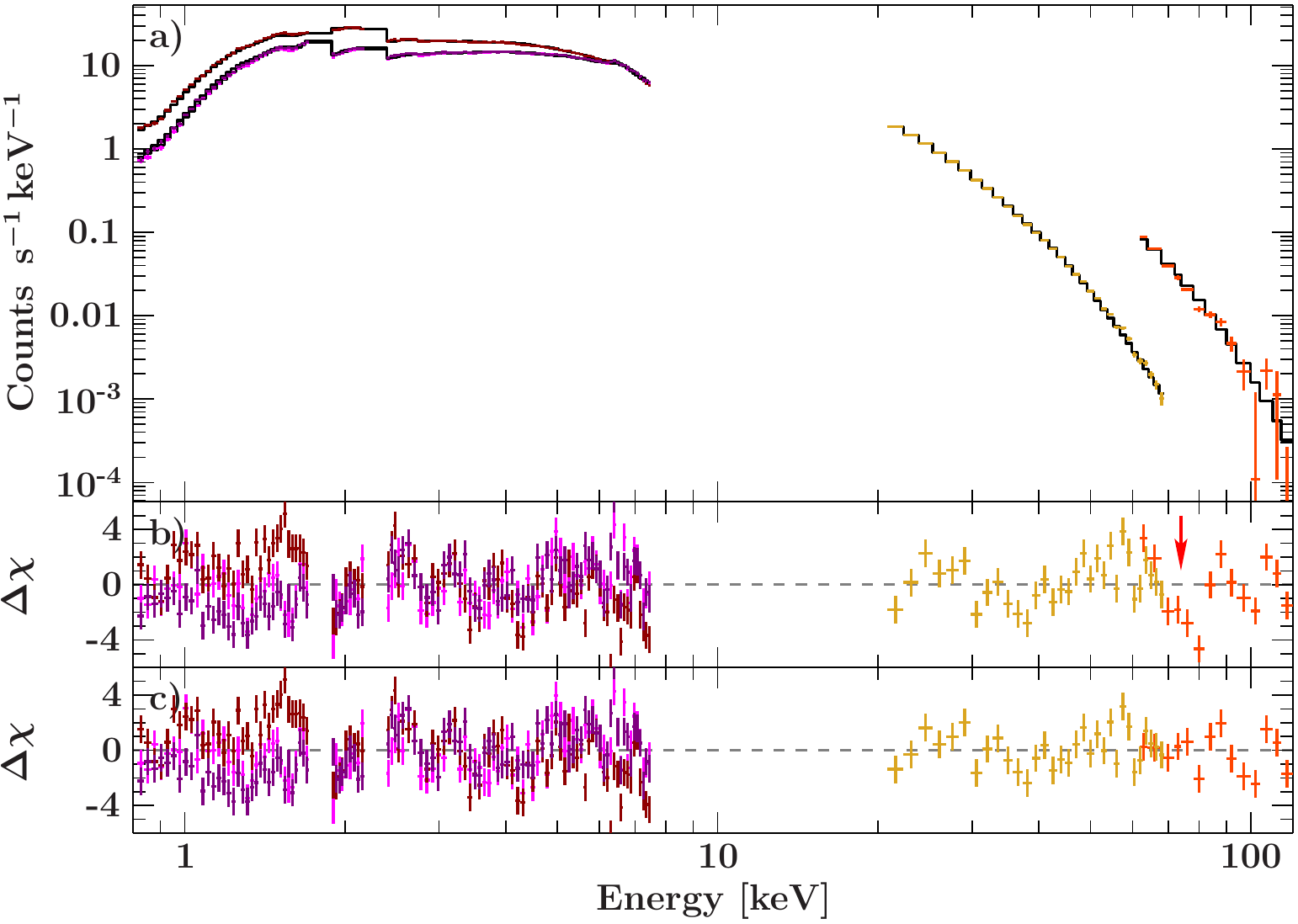}
\caption{
\suzakun-only fit.  XIS data are red, pink, and purple, PIN data are
yellow, and GSO data are orange.  Panels as in Figure \ref{fig:spectra_n}.
\label{fig:spectra_s}}
\end{figure}

\begin{figure}
\includegraphics[width=\columnwidth]{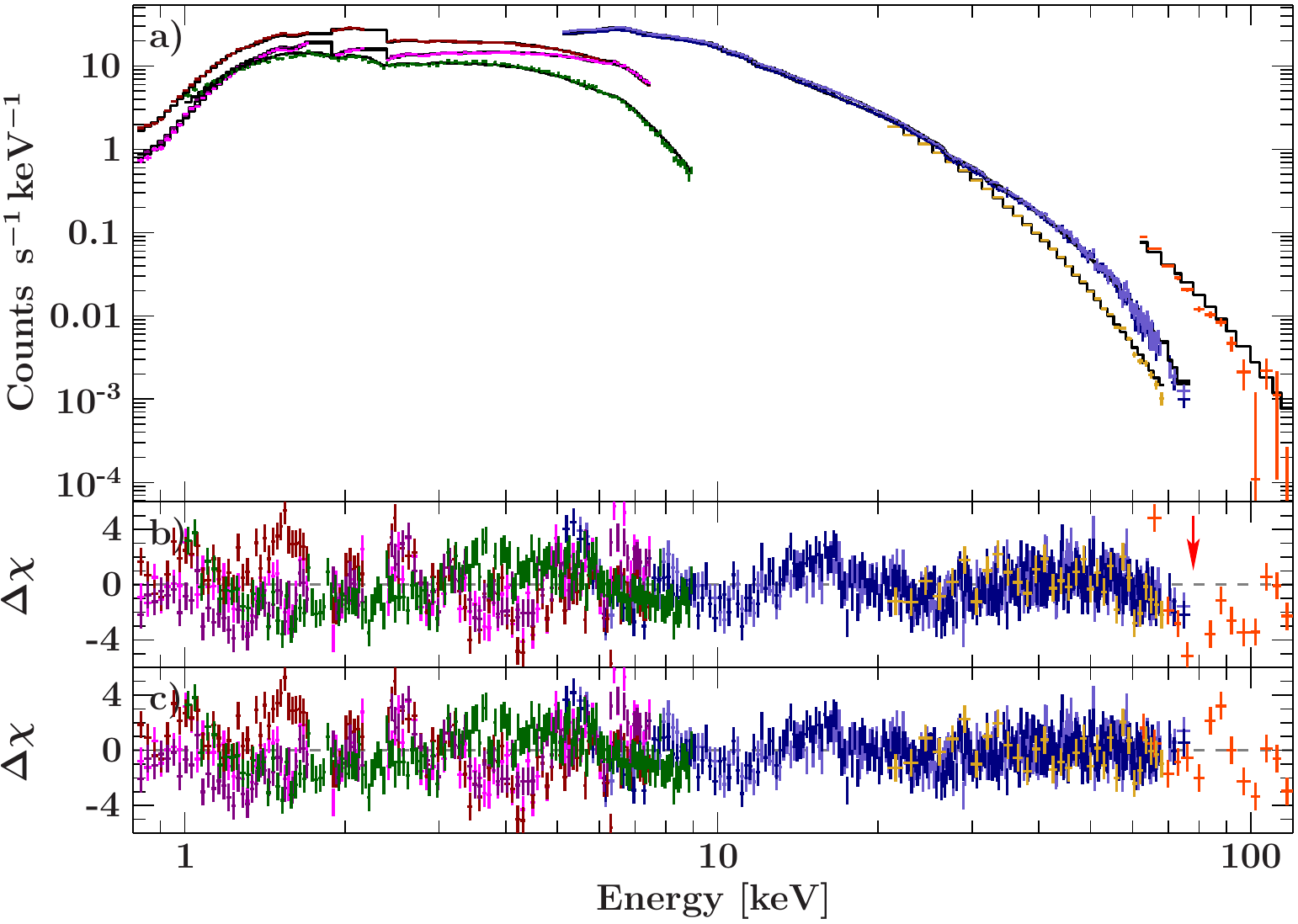}
\caption{
Joint \nustarn--\swiftn--\suzaku fit.  Panels as in Figure \ref{fig:spectra_n}.
\label{fig:spectra_nws}}
\end{figure}

In contrast to the fits to Type~I bursts reported by
\citet{Kuhnel:2013:GROJ1008},
we found the \texttt{npex} model provides a better fit for all combinations
of instruments despite having fewer free parameters, so we restrict our
attention to this model for further analysis.
\citet{tmp_Yamamoto:14:GROCyc} similarly found that the \texttt{npex} model
provided the best fit to the \suzaku data from this giant outburst.
Table \ref{tab:npex_fits}
provides the best-fit values for the time-integrated continuum parameters, and
Figures \ref{fig:spectra_n}--\ref{fig:spectra_nws} 
show the best fits. 

The brightness of the source highlights systematic effects in joint fits
between multiple instruments, producing poor goodness of fit.  Only the
\nustarn-only fit has a reasonable goodness of fit, at $\chi^2_\nu = 1.18$
for 536 degrees of freedom.
There is substantial disagreement
between the instruments below 10\,keV (e.g., Figure \ref{fig:spectra_nws}).
The \nustar and \suzaku observations are not simultaneous, and so some
spectral evolution may have occured between the two epochs. However, there
is also
disagreement even among the three XIS modules (Figure \ref{fig:spectra_s}).
This disagreement at low energies, driven primarily by XIS1,
leads to differences in the best-fit blackbody
temperature and power-law indices (Table \ref{tab:npex_fits}).
These discrepancies were also noted in this dataset 
by \citet{Kuhnel:2013:GROJ1008} and \citet{tmp_Yamamoto:14:GROCyc}; these
authors elected to excise several energy ranges from the XIS
backside-illuminated detectors or exclude the data entirely.
The fit for the blackbody temperature shows multiple minima 
in the \suzakun-only
fit, for example, with the $\sim3$\,keV temperature preferred to the
$0.4$\,keV temperature suggested by the joint fit to all instruments.
Similarly, our coarser binning and inclusion of all three XIS modules 
results in better fits to the iron line complex with two broadened
Gaussians rather than the narrow 6.4, 6.67, and 7\,keV lines fit by
\citet{Kuhnel:2013:GROJ1008} and \citet{tmp_Yamamoto:14:GROCyc}.
We are primarily interested in the
spectral behavior at high energies, which is well above the folding energy
$E_{\rm fold}$ and hence relatively insensitive to these parameters.

\subsection{Evidence for a Cyclotron Line} \label{sec:cyc_line}

Next, we fit the data using the above 
continuum model and a multiplicative cyclotron
scattering feature using a pseudo-Lorentzian optical depth profile
(the XSPEC \texttt{cyclabs} model, \citealp{Mihara:90:cyclabs}).  
We initially confined
our search to line centers above 50\,keV, with fits initialized near the
75\,keV value reported by \citet{Yamamoto:2012:GROSuzakuCyc}.
Table \ref{tab:npex88_fits} reports the best-fit parameters, while Figures
\ref{fig:spectra_n}--\ref{fig:spectra_nws} compare
the residuals to fits without a cyclotron line.

Fits to the \nustar data alone do not provide strong constraints on the
CRSF parameters,  as there are degeneracies with the continuum modeling  
because the cyclotron line lies at the upper edge of the \nustar bandpass.
However, there are clear residuals in the \nustar data above 70\,keV, and
\nustarn-only fits are significantly improved by the CRSF.
The best-fit \suzaku CRSF parameters are a reasonable match to those reported in
\citet{tmp_Yamamoto:14:GROCyc} given the minor differences in analysis
methods.

Combining the \nustar data with \suzaku provides an independent confirmation
of the line.
In the joint fit, the line centroid moves to 78\,keV and the best-fit
width is 11\,keV, matching the values obtained by
\citet{tmp_Yamamoto:14:GROCyc} in their \texttt{npex} fits including XIS.
The GSO
cross-normalization changes from 1.19 relative to \nustar in the
\texttt{npex} fit to 1.38 in the \texttt{npex} with cyclotron feature fit.
All other cross-normalization constants remain constant within errors
(Table \ref{tab:norms}).  The cyclotron-line fit thus produces correctly
the expected agreement between the normalizations of \suzaku PIN and GSO.

\input{table_norms}

Both the \nustar and \suzaku data thus independently show evidence for a
CRSF in the 70--80\,keV range.
Because the \nustar data do not cover the
entire CRSF, the joint fit provides the best constraint on the line
parameters, but the parameters are sensitive to the \nustar and \suzakun-HXD
cross-calibration.

We assessed the significance of the detections using the method of
posterior predictive p-values \citep[ppp-values,][]{Protassov:02:LRTest}.  
Briefly, we simulate many
instances of a model without a \texttt{cyclabs} feature
by folding the spectral model through instrumental
responses; the exposure and binning are matched to the real data, and the
data and background are perturbed to account for counting statistics.
For each simulated dataset, we 
then fit the null model and a test model with a \texttt{cyclabs} feature.
For each simulated realization, we determine
$\Delta \chi^2$ between the two models and
compare the distribution of $\Delta \chi^2$ values for the simulations 
to the observed value. 

If few of the simulated $\Delta \chi^2$ values are as large as observed in
the real data, this provides evidence for the CRSF.
Rather than restricting the simulated model parameters to those of the
best-fit null model, we use the Cholesky method to 
draw the simulated parameters from a multivariate
Gaussian derived from the covariance matrix obtained in the fit to the null
model \citep{Hurkett:08:XRTLineSearches}.

We performed 10,000 simulations of the \nustar data alone as well as the
joint \nustarn, \swiftn, and \suzaku data.  The line energy was allowed to
vary in the 50--100\,keV range and the line width between 1 and 30\,keV.  In all
cases, the simulated $\Delta \chi^2$ was less than the value observed in
the real data, providing $>3.9\sigma$ evidence for the existence of
the line in each of the two fits.  
In most of the simulated cases, the best-fit depth of the line
is zero, and so the two models are indistinguishable.  The largest
deviation in $\chi^2$ was 17.0 (21.3) for the combined datasets (\nustar
only), far smaller than the $\Delta \chi^2$ values of 278.5 (62.8) seen in
the real data.  Based on the difference between the observed and simulated
$\Delta \chi^2$ distributions, it is clear that the CRSF detection is much
more significant in the joint fit than when using \nustar alone.

Given the distribution of $\Delta \chi^2$ in these simulations,
it would be computationally unfeasible to simulate enough realizations to 
expect a $\Delta \chi^2$ value near the true value and 
obtain a true ppp-value significance.  
We can obtain a simple estimate of the significance (and hence the number of 
simulations required to obtain that chance deviation) by summing the data
and model counts in the $\pm1\sigma$ energy window around the best-fit
78\,keV cyclotron line.  
Dividing the difference between the \texttt{npex} model without the cyclotron line
and the data in this region by the statistical error 
allows us to estimate the level of chance
fluctuation needed.  The deviations in the 
\nustar data (which do not cover the full
cyclotron line) are 1.8$\sigma$ and 2.3$\sigma$ when the modules are
considered independently; 
the deviation in the GSO data taken alone is 8.0$\sigma$.
We thus expect the GSO measurement to dominate the fit.
(We do not correct for trials over energy 
because the high-energy line was previously reported in other observations.)
If taken at face value, the statistical errors would require more than
$8\times10^{14}$ simulations to achieve deviations in $\chi^2$ comparable to the
observed values.

We considered whether systematic calibration uncertainties could be
responsible for the observed feature.
While the method of ppp-values provides a robust assessment of line
significance \citep{Hurkett:08:XRTLineSearches}, it is sensitive to false
positives if systematic errors are present.  If a line feature is due to
inaccuracy in the instrumental responses or the modeled background,
ppp-value tests will confirm its statistical significance but not its
physical reality.   The calibration of the \nustar responses in the
70--78\,keV range is less certain than at lower energies due to the
increasing faintness of astrophysical calibrators.  However, measured deviations
from a fiducial spectrum of the Crab Nebula are $<15$\% from 70--78\,keV
(Madsen et al. 2014, in prep).
Similarly, few-percent deviations of the Crab spectrum have been measured in 
\suzaku GSO spectra near 70\,keV \citep{Yamada:11:GSOCal}.  These effects
are not large enough to produce the $\sim$30\% deviation 
seen here, so we conclude that the feature is both significant and real.

\subsection{Search for a Lower-Energy Fundamental Line}
\label{sec:fundamental_search}

We searched for a cyclotron line at half the energy of the 78\,keV line
reported in Section \ref{sec:cyc_line}.  
The \nustar data enable a more sensitive search than
is possible with PIN: The combined \nustar data have an SNR of 135 per keV at
40\,keV in these data, compared to 60 for PIN.  (Data from both instruments
are strongly source-dominated, given the brightness of the outburst.)
No obvious residuals are apparent
in the time-integrated \nustar data near $\sim$38\,keV (Figure
\ref{fig:spectra_n}), consistent with previous non-detections in
time-integrated data.  Some residual structure is present in the
\suzakun-PIN data below 40\,keV (Figure \ref{fig:spectra_s}).  Using PIN
data, \citet{tmp_Yamamoto:14:GROCyc} reported a possible fundamental with
$E_0 = 36.8\spm{1.1}{0.7}$\,keV, optical depth at line center 
of $0.06\spm{0.08}{0.03}$, and width
of $11.1\spm{7.2}{10.2}$\,keV in their \suzakun-only fit, but concluded it
is not statistically significant.  A double-cyclotron line
\nustarn--\suzakun--\swift joint fit with the fundamental restricted to
half of the 90\% error limits for the 78\,keV does fit a line depth at
39.8\spm{0.5}{1.2}\,keV.  It has depth 1.0\spm{0.7}{0.2} and width
6.0\spm{8.4}{4.5}\,keV.  However, the improvement in $\chi^2$ is modest,
only 6.3 for three additional free parameters.  Our ppp-value simulations
of the higher energy line found 47 out of 10,000 simulations had $\Delta
\chi^2$ values larger than this, implying $<2.9\sigma$ significance.  A
\nustarn-only fit to a line at this position results in a fundamental with
depth consistent with zero.  The 90\% CL upper limit on the optical depth
at 39\,keV is 0.04.  The possible 39\,keV fundamental fit by the \suzaku
data is thus disfavored by the more sensitive \nustar data.  A broader
\nustar search from 34--40\,keV (at half of the line centroid identified by
the independent \nustar and \suzaku fits) similarly yielded line depths
consistent with zero.

We also performed a generic search for lower-energy lines by stepping a
\texttt{cyclabs} feature through a 2\,keV grid of energies over the 
10--60\,keV range.  
We used the \nustar data only, as in the joint fit the residuals show
dips (due to response differences highlighted by the brightness of the
source, Figure \ref{fig:spectra_nws}c) not present in the \nustarn-only 
fit (Figure \ref{fig:spectra_n}c).
For speed, the continuum parameters were frozen in the initial search.
Only one trial (at 26\,keV) fit a line depth greater than zero, and in
this case the best fit line width was 1\,keV, at the narrowest limit.
The value of  $\Delta \chi^2$ is 5.4, less than the largest values 
obtained by chance coincidence in the Monte Carlo simulations of the
80\,keV line, and there is a known response calibration feature at this
energy.
Over all energies, the largest 90\% CL upper limit on the optical depth was 
0.09 at 52\,keV.  Accordingly, we can rule out a lower-energy fundamental with
greater depths in the time-integrated data.

\subsection{Phase-Resolved Fits}
\label{sec:phase_resolved}

Because cyclotron line intensities and energies may vary with pulse phase
\citep[e.g.,][]{Fuerst:13:HerX1},
we split the \nustar observation into phase bins of roughly
constant signal-to-noise ratio and conducted spectral fits on each.

\begin{figure}
\includegraphics[width=\columnwidth]{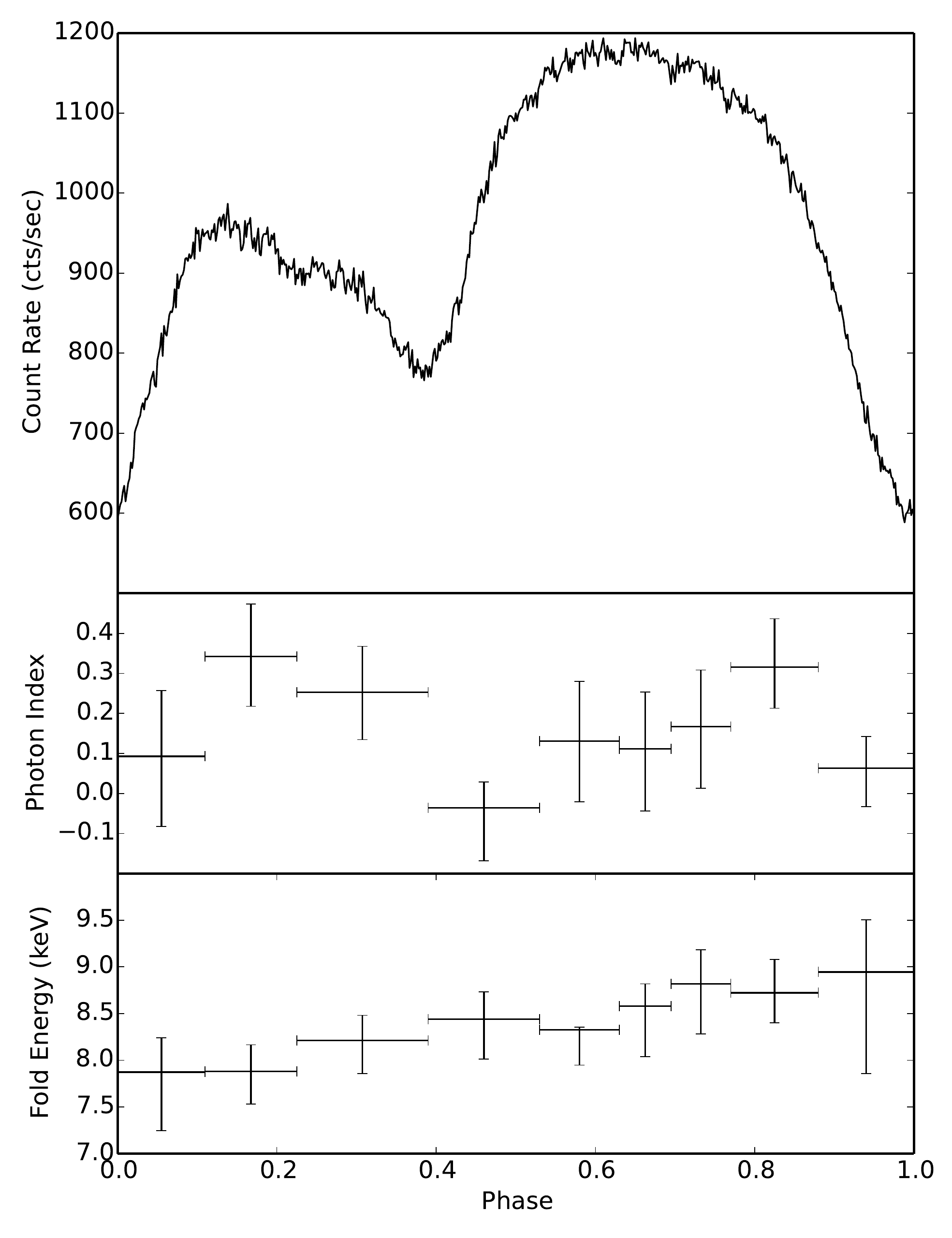}
\caption{\nustar data (3--79\,keV) folded at the 93.57\,sec spin period
(top panel)
and the best-fit phase-resolved \texttt{npex} photon indices (middle
panel) and folding energies (bottom panel).
\label{fig:phase_folded}
}
\end{figure}

We barycentered the \nustar event data with \texttt{barycorr} using the
DE200 solar system reference frame \citep{Standish:82:DE200Ephemeris} and
applied a correction for light-travel time across the binary orbit using the
ephemeris of \citet{Kuhnel:2013:GROJ1008}.  Figure \ref{fig:phase_folded}
shows the \nustar data folded at the best-fit spin period of 93.57434\,sec.  

The time-resolved data were well-fit by the \texttt{npex} spectral model.
We fixed the positive power-law index to 2, consistent with the values
obtained in the time-integrated fits.
The photon indices show a correlation with intensity (Figure
\ref{fig:phase_folded}), while the folding energy shows a mild secular
increase throughout the pulse.
No obvious deficits are present in the residuals at lower energies.

We attempted to observe phase evolution of the 80\,keV CRSF.  Using four
phase bins, \citet{tmp_Yamamoto:14:GROCyc} found only a slight dependence
of the CRSF energy on the pulse data using the \suzaku data.  We froze the
width and depth of the CRSF to the best-fit values from the joint
\nustarn--\suzakun--\swift fit but left the energy free to vary.  However,
the relatively limited \nustar energy coverage of the line does not permit
firm constraints on the line energy.  

We also performed a harmonic CRSF fit to the phase-resolved data.  We fit
for a fundamental line in the 34--47\,keV range and froze the $\sim$80\,keV
harmonic width and depth to the best-fit time-integrated values.  In all
epochs, the line depth was consistent with zero, the linewidth
unconstrained, and the improvement in $\chi^2 < 2$.  
The 90\% C.L. line depth upper limits were in the range 0.09--0.3.

We repeated the generic grid search for low-energy CRSFs in the
time-resolved spectra in case phase-dependent intensity was pushing a
fundamental below detectability in the time-integrated fit.  As in the
time-integrated case, the additional CRSF fit widths pegged at the
minimum value of 1\,keV, fit depths that were consistent with zero, 
and/or were associated
with the small known response feature at 26\,keV.
Over all energies, the largest upper limits on the line depth were
0.2--0.4 and occurred in the 50--60\,keV range.

Accordingly, our phase-resolved fits rule out a phase-dependent fundamental
CRSF below 70\,keV with depth greater than one third of the depth of the
80\,keV CRSF.  

\section{Discussion}  \label{sec:discussion}

Observations of the November 2012 giant outburst of \gro with modern
instruments provide the best available constraints on the magnetic field
strength of this HMXB.
Our spectral fits have confirmed that the previously reported CRSF
is indeed the fundamental for \gron.  The best-fit line center for the
combined datasets is 
78\spm{3}{2}\,keV.  This
matches the CRSF reported by \citet{tmp_Wang:14:GROJ1008_INTEGRAL} using
\textit{INTEGRAL} data 
but is lower than the 88$\pm$5\,keV value first reported by
\citet{Grove:95:GROJ1008CRSF}.
This difference is not highly significant, however. 
The \suzaku data provide a better
constraint on the line and higher significance detection because the line
centroid is at the upper edge of the \nustar bandpass, 
but the \nustar data provide an
independent confirmation of the detection.  

The higher sensitivity provided by \nustar below 79\,keV enabled us to
perform the most constraining search for a fundamental CRSF at lower
energies.  
Our \nustar double-cyclotron line fits require
the ratio of the optical depths of the fundamental to the harmonic to be
less than 5\% in the time-integrated data.
This is less than the most extreme ratios observed 
in other accreting pulsars, including Vela X-1 ($\sim$10\%;
e.g., \citealp{Fuerst:14:VelaX1}) and 4U~0115$+$634 ($\geq11\%$,
e.g., \citealp{Muller:13:4U0115}).  
In both of those systems, however, phase-resolved fitting reveals intervals
of greater fundamental strength.  While our phase-resolved limits on the
fundamental/harmonic optical depth ratios are less stringent ($<$11--37\%) than the
time-integrated constraint, we do not
detect a significant fundamental at any phase.

Photon spawning can weaken the strength of the observed fundamental:
an electron that scatters into an excited Landau state 
will release one or more secondary photons with energy comparable to the
line energy that may escape to the observer.
Calculations suggest this process can
replace as much as 75\% of the flux in the fundamental CRSF
\citep{Schonherr:07:CRSF}.  
It thus is difficult to account for the low time-integrated fundamental to
harmonic depth ratio we observe with spawning alone.
Moreover, spawning is influenced by the hardness
of the spectral continuum, with harder spectra producing weaker fundamental
lines with more pronounced emission wings \citep{Schonherr:07:CRSF}.
Our nondetection of a low-energy fundamental in
the time-resolved fits despite the phase variation in the continuum
spectrum (Figure \ref{fig:phase_folded}) thus argues against such masking.

We therefore conclude that the 78\,keV CRSF is likely the fundamental.
The inferred magnetic field strength for \gro is 
$6.7\times10^{12} (1+z)$\,G, the highest of known accreting pulsars.

\acknowledgments
This work was supported under NASA Contract No. NNG08FD60C and uses
data from the \nustar mission, a project led by the California Institute of
Technology, managed by the Jet Propulsion Laboratory, and funded by the
National Aeronautics and Space Administration. We thank the \nustar
Operations team for executing the ToO observation and the Software and 
Calibration teams for analysis support.
This research has used the 
\nustar Data Analysis Software (NuSTARDAS) jointly developed by the ASI
Science Data Center (ASDC, Italy) and the California Institute of
Technology (USA).
JW acknowledges partial support from Deutsches Zentrum f\"ur Luftund Raumfahrt
grant 50\,OR\,1113.

{\it Facilities:} \facility{NuSTAR}, \facility{Swift}, \facility{Suzaku}.


\bibliographystyle{yahapj}
\bibliography{nustar}

\clearpage
\pagebreak[4]
\clearpage
\input{table_npexplus.tex}
\input{table_npex88.tex}

\end{document}

%% file: table_norms.tex
\begin{table}
\centering
\caption{Best-fit cross-normalization constants relative to \nustar 
Module A for the joint
\nustarn--\swift--\suzaku fit including a cyclotron line.
Errors are 90\% C.L. 
\label{tab:norms}
}
\begin{tabular}{ll} 
 \hline 
Instrument & Normalization \\
\hline  
\nustar A & $\equiv1.0$ \\
\nustar B & 1.014$\pm${0.001} \\
\swift XRT &	1.325$\pm${0.007} \\
\suzaku XIS0 &	1.074$\pm${0.003} \\
\suzaku XIS1 &	1.095$\pm${0.003} \\
\suzaku XIS2 &	1.115$\pm${0.003} \\
\suzaku PIN &	1.380$\pm${0.004}  \\
\suzaku GSO &	1.39$\pm${0.04}  \\
\hline 
\end{tabular}
\end{table}

%% file: table_npexplus.tex
\begin{table}
\centering
\caption{Best-fit time-integrated \texttt{npex} parameters for fits with 
\nustar (N), \nustar and \swift (NX), \suzaku (S), and all instruments
(NXS).  Errors are 90\% C.L.  In all fits the positive powerlaw index
$\Gamma_2$ was free to vary but converged to the limiting value of 2.
Normalization units are photons\,keV$^{-1}$\,cm$^{-2}$\,s$^{-1}$ 
for the powerlaw
components ($A_{1,2}$), photons\,cm$^{-2}$\,s$^{-1}$ for the Gaussians
($A_{\rm gauss 1,2}$), and $L_{39}/D_{10}^2$ for the black body ($A_{\rm BB}$),
where $L_{39}$ is the source luminosity in units of $10^{39}$ ergs$^{-1}$
and $D_{10}$ is the distance to the source in units of 10\,kpc.
\label{tab:npex_fits}
}
\begin{tabular*}{0.55\textwidth}{lllll} 
 \hline 
Parameter  & N & NX  & S  & NXS  \\ 
\hline  
$N_{\rm H}$ [$10^{22}$ cm$^{-2}$] &   $1.1^{+1.2}_{-1.1}$   &   $1.06\pm0.06$   &
$1.60\pm0.03$   &   $1.45\pm0.02$    \\ 
\hline
$\Gamma_1$   &   $0.15\pm0.05$   &   $0.27^{+0.06}_{-0.04}$ &
$0.94^{+0.02}_{-0.05}$   &   $0.30\pm0.01$    \\ 
$A_1$  [$10^{-1}$]    &   $1.76^{+0.26}_{-0.20}$   &   $2.38\pm0.15$    &
$5.17^{+0.05}_{-0.15}$   &   $2.70^{+0.08}_{-0.10}$    \\ 
$\Gamma_2$   &  2 & 2 & 2 & 2 \\ 
$A_2$  [$10^{-4}$]    &   $1.39^{+0.30}_{-0.18}$   & $2.0^{+1.6}_{-0.4}$    &   $4.7^{+2.0}_{-0.4}$   &   $1.81\pm0.07$    \\ 
$E_{\rm fold}$ [keV] &   $8.44^{+0.14}_{-0.13}$   &   $8.34^{+0.38}_{-0.17}$   &   $7.47^{+0.16}_{-0.08}$   &   $8.17\pm0.05$    \\ 
\hline  
$E_{0,1}$  [keV]   &   $6.60\pm0.02$   &   $6.60\pm0.02$   &    $6.59\pm0.02$   &   $6.58\pm0.01$    \\ 
$\sigma_1$  [keV]   &   $0.29\pm0.02$ &   $0.30\pm0.02$   & $0.28\pm0.03$   &   $0.27\pm0.02$    \\ 
$A_\mathrm{gauss,1}$ [$10^{-3}$]  &   $6.8\pm0.7$   &   $7.1\pm0.4$   & $9.0^{+1.3}_{-1.4}$   &   $6.5\pm0.4$    \\ 
$E_{0,2}$  [keV]   &   &  &  $5.4^{+0.2}_{-0.1}$   &   $6.0^{+0.2}_{-0.3}$    \\ 
$\sigma_2$  [keV]   &   &   &   $0.07^{+0.03}_{-0.02}$   &   $0.19\pm0.02$    \\ 
$A_\mathrm{gauss,2}$ [$10^{-3}$]   &   & &  $7.4^{+0.5}_{-0.2}$   &   $27^{+7}_{-5}$    \\ 
\hline  
$kT$  [keV]   &   $1.61\pm0.03$   &   $1.62^{+0.09}_{-0.07}$   &
$3.4^{+0.3}_{-0.2}$   &   $0.42\pm0.02$    \\ 
$A_\mathrm{BB}$ [$10^{-3}$]   &   $14\pm4$   &   $8.7^{+1.7}_{-1.4}$   &
$79^{+5}_{-7}$   &   $3.9^{+0.3}_{-0.2}$    \\ 
\hline  
$\chi^{2}/\mathrm{dof}$ &  632.6/536&  881.8/692&  1134.8/343&  2424.9/1045   \\ 
$\chi^{2}_\mathrm{red}$ &  1.18&  1.27&  3.31&  2.32   \\ 
\hline 
\end{tabular*}
\end{table}

%% file: table_npex88.tex
\begin{table}
\centering
\caption{
Best-fit time-integrated \texttt{npex} parameters with a cyclotron feature
for fits with
\suzaku (S) and all instruments
(NXS).  Errors are 90\% C.L.  Normalizations units are as in Table
\ref{tab:npex_fits}.
\label{tab:npex88_fits}
}
\begin{tabular*}{0.4\textwidth}{lll} 
 \hline 
Parameter  & S  &  NXS  \\ 
 \hline 
$N_{\rm H}$ [$10^{22}$ cm$^{-2}$] &      $1.62^{+0.04}_{-0.06}$   &   $1.43^{+0.03}_{-0.02}$    \\ 
\hline  
$E_\mathrm{0, cyc}$ [keV] &     $74^{+3}_{-2}$   &   $78^{+3}_{-2}$    \\ 
\texttt{cyclabs} width [keV] &   $<5.89$   &   $11^{+6}_{-4}$    \\ 
\texttt{cyclabs} depth &     $4.2^{+3.8}_{-3.3}$   &   $0.81^{+0.15}_{-0.13}$    \\ 
\hline  
$\Gamma_1$   &     $1.03^{+0.12}_{-0.13}$   &   $0.29^{+0.03}_{-0.02}$    \\ 
$A_1$  [$10^{-1}$]    &    $5.24^{+0.23}_{-0.29}$   &   $2.54^{+0.13}_{-0.17}$    \\ 
$\Gamma_2$   &    $1.10^{+0.04}_{-0.20}$   &   2 \\
$A_2$  [$10^{-4}$]    &   $52.479^{+0.005}_{-7.573}$   &   $1.40^{+0.29}_{-0.19}$    \\ 
$E_{\rm fold}$ [keV] &    $8.9\pm1.0$   &   $8.55^{+0.31}_{-0.14}$    \\ 
\hline  
$E_{0,1}$  [keV]   &     $6.589^{+0.002}_{-0.010}$   &   $6.58^{+0.01}_{-1.44}$    \\ 
$\sigma_1$  [keV]   &    $0.29\pm0.03$   &   $0.27\pm0.02$    \\ 
$A_\mathrm{gauss,1}$ [$10^{-3}$]  &      $9.3^{+1.3}_{-1.2}$   &   $6.5\pm0.4$    \\ 
$E_{0,2}$  [keV]   &    $5.36^{+0.17}_{-0.10}$   &   $5.6^{+0.4}_{-0.5}$    \\ 
$\sigma_2$  [keV]   &   $0.63^{+0.25}_{-0.14}$   &   $2.07^{+0.32}_{-0.27}$    \\ 
$A_\mathrm{gauss,2}$ [$10^{-3}$]   & $5.7^{+0.8}_{-1.9}$   &   $34^{+16}_{-9}$    \\ 
\hline  
$kT$  [keV]   &    $2.90^{+0.99}_{-0.26}$   &   $0.45\pm0.02$    \\ 
$A_\mathrm{BB}$ [$10^{-3}$]  &     $43^{+31}_{-20}$
&   $4.2^{+0.5}_{-0.4}$    \\ 
\hline  
$\chi^{2}/\mathrm{dof}$ &  1070.7/340 &  2266.0/1042   \\ 
$\chi^{2}_\mathrm{red}$ &  3.15 &  2.17   \\ 
\hline
\end{tabular*}
\end{table}